\documentclass{cs23proc}

\usepackage{kantlipsum}
\usepackage{url}
\usepackage{xcolor}

\editors{Takeru Suzuki and the Cool Stars 23 Organizing Team}
\publisher{Zenodo}
\conference{The 23th Cambridge Workshop on Cool Stars, Stellar Systems, and the Sun (Cool Stars 23)}
\conferencedate{2026}

\title{Magnetic Heating Across the Sun and Solar-like Stars: Universal Scaling Laws from Chromospheres to Coronae}
\author{SHIN TORIUMI $^{1}$,
        KOSUKE NAMEKATA $^{2,3}$,
        YUTA NOTSU $^{4,5}$,
        MAI YAMASHITA $^{6}$,
        VLADIMIR S. AIRAPETIAN $^{3,7}$}

\affiliation{$^{1}$ Institute of Space and Astronautical Science, Japan Aerospace Exploration Agency, 3-1-1 Yoshinodai, Chuo-ku, Sagamihara, Kanagawa 252-5210, Japan \\
			 $^{2}$ The Hakubi Center for Advanced Research, Kyoto University, Yoshida-Honmachi, Sakyo-ku, Kyoto, 606-8501, Kyoto, Japan \\
			 $^{3}$ NASA Goddard Space Flight Center, 8800 Greenbelt Road, Greenbelt, 20771, MD, USA \\
			 $^{4}$ Laboratory for Atmospheric and Space Physics, University of Colorado Boulder, 3665 Discovery Drive, Boulder, 80303, CO, USA \\
			 $^{5}$ National Solar Observatory, 3665 Discovery Drive, Boulder, 80303, CO, USA \\
			 $^{6}$ Department of Applied Physics, Interdisciplinary Faculty of Science and Engineering, Shimane University, 1060 Nishikawatsu, Matsue, Shimane, 690-8504, Japan \\
			 $^{7}$  American University, 4400 Massachusetts Avenue NW, Washington, DC 20016 USA \\ }

\shorttitle{Magnetic Heating Across the Sun and Cool Stars}
\shortauthors{TORIUMI et al.}

\abs{
Magnetic activity in cool stars governs the thermal structure and high-energy radiative output of their outer atmospheres, thereby influencing stellar evolution, stellar winds, and the atmospheres of orbiting (exo)planets. A long-standing question in stellar astrophysics is whether the mechanisms responsible for atmospheric heating are universal across the Sun and cool stars spanning different ages and activity levels. In this paper, we review recent progress in understanding magnetic heating across the Sun and Sun-like stars through empirical scaling relations between photospheric magnetic flux and radiative output from the chromosphere, transition region, and corona. Analysis of more than a decade of Sun-as-a-star observations reveals that irradiance and magnetic flux follow power-law relationships over a wide temperature range. While coronal emissions exhibit superlinear scaling with magnetic flux, chromospheric and transition-region diagnostics show weaker, sublinear dependencies. Remarkably, observations of G-type stars with ages ranging from 50 Myr to 4.5 Gyr are found to lie on extensions of the solar scaling laws, suggesting that a common magnetic-heating mechanism operates across different levels of stellar activity. We further discuss the application of these scaling relations to reconstructing stellar X-ray and ultraviolet (XUV) spectra from observed magnetic fluxes. The resulting synthetic spectra reproduce actual observations of young, active solar analogs, providing a practical tool to estimate the ionizing radiation whose extreme UV emissions cannot be directly measured.
}

\begin{document}

\maketitle

\section{Introduction}

Cool stars with outer convective envelopes, including F-, G-, K-, and M-type stars, generate magnetic fields through dynamo action operating within their interiors \citep{1984ApJ...279..763N}. Magnetic fields emerging at the stellar surface transport and dissipate energy in the outer atmosphere, thereby heating the chromosphere, transition region, and corona to temperatures far exceeding that of the photosphere \citep{2004A&ARv..12...71G,2006SoPh..234...41K}. As a consequence, these hot atmospheres produce intense X-ray and ultraviolet (UV) radiation and drive stellar winds that carry mass and angular momentum away from the star \citep{2006JGRA..111.6101S,2014MNRAS.441.2361V,2020IJAsB..19..136A}. Such high-energy radiation and outflowing winds strongly influence the environments of orbiting (exo)planets and are therefore key ingredients in understanding the evolution and habitability of planetary atmospheres \citep{2016NatGe...9..452A,2019LNP...955.....L}. Magnetic activity also manifests itself through the formation of starspots and the occurrence of flares and eruptions, which are among the most prominent signatures of magnetic energy storage and release in stellar atmospheres \citep{2011LRSP....8....6S,2017LRSP...14....4B,2017ApJ...834...56T,2019LRSP...16....3T}.

Stellar magnetic activity evolves with stellar rotation and age. Young rapidly rotating stars generate strong magnetic fields through efficient dynamo action and consequently exhibit enhanced X-ray and ultraviolet (XUV) emissions. Over time, angular momentum loss through stellar winds spins down the star, reducing its magnetic activity and high-energy radiative output \citep{1972ApJ...171..565S,2011ApJ...743...48W}. Recent observations have revealed that some young and active Sun-like stars produce superflares with energies 10 to 1000 times greater than the largest solar flares, indicating that magnetic energy release in cool stars can far exceed that observed on the present-day Sun \citep{2012Natur.485..478M,2013ApJ...771..127N}. In some cases, associated eruptive filaments have also been detected, suggesting that highly active stars may experience extreme space-weather events \citep{2022NatAs...6..241N}.

Consequently, the enhanced magnetic activity of young stars implies a substantially stronger XUV radiation environment than that of the present-day Sun \citep{2005ApJ...622..680R}. This enhanced high-energy radiation can drive vigorous photochemistry in planetary atmospheres and promote atmospheric escape, thereby influencing the long-term evolution and habitability of orbiting planets.

\begin{figure*}[ht!]
	\centering
	\includegraphics[width=180mm]{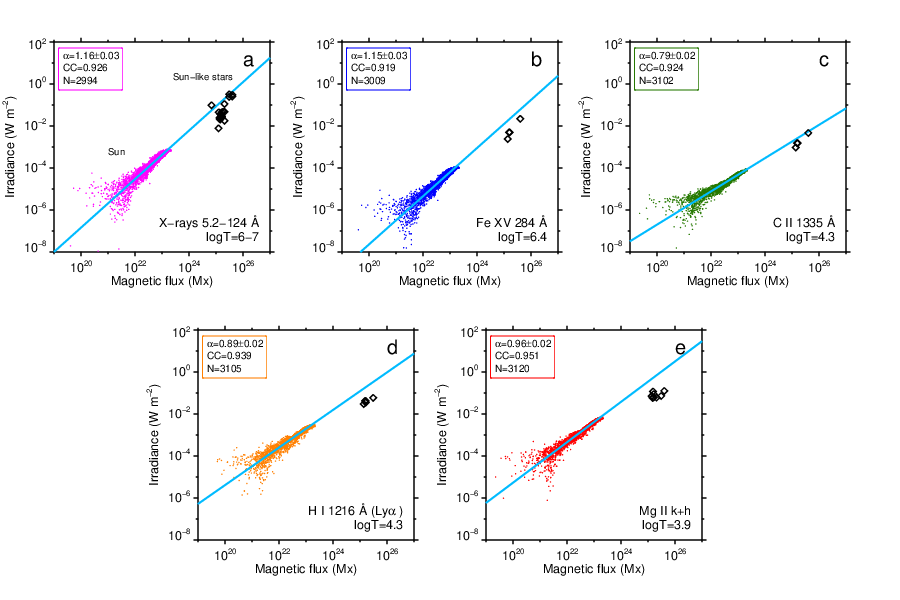}
	\caption{Magnetic flux-irradiance relationships for representative atmospheric diagnostics. Each panel shows the power-law index $\alpha$, correlation coefficient $CC$, and number of data points $N$ for the solar data. Diamonds indicate the stellar data from the literature. Power-law correlations are found from the chromosphere to the corona, although the power-law index $\alpha$ varies with atmospheric temperature. Figure is reproduced from \citet{2022ApJ...927..179T}; however, the stellar magnetic flux for the stars was recalculated by correcting the error in the filling factor.}
	\label{fig:cc}
\end{figure*}

Understanding the origin of stellar XUV radiation therefore requires an understanding of how stellar atmospheres are heated. A fundamental question is whether the Sun and cool stars share a common atmospheric heating mechanism. Previous studies have demonstrated that magnetic flux and coronal X-ray emission follow a universal power-law relation over a wide range of magnetic activity levels \citep{1998ApJ...508..885F,2003ApJ...598.1387P}. Building on these findings, recent solar-stellar studies have extended the investigation from the corona to cooler atmospheric layers and revealed empirical scaling laws linking magnetic flux to radiative output from the chromosphere, transition region, and corona \citep{2022ApJ...927..179T,2022ApJS..262...46T}. A recent review by \citet{2026SoPh..301...48T} summarized these developments in the broader context of the solar-stellar connection and highlighted the role of Sun-as-a-star observations in bridging solar and stellar physics. In this paper, we review the universal scaling laws and discuss their application to reconstructing the XUV spectra of active Sun-like stars \citep{2023ApJ...945..147N}.
%Now, for something really huge. I mean, really huge. Like, huger than huge can be. We're going to end a sentence with a footnote.\footnote{ This is what a footnote looks like.}

\section{Magnetic Flux-Irradiance Scaling}\label{sec:scaling}

\subsection{Historical X-ray Scaling Laws}

The relationship between the photospheric magnetic flux and coronal X-ray emission has long served as a fundamental observational constraint on the atmospheric heating of the Sun and stars. Using a wide variety of solar and stellar magnetic structures, ranging from quiet-Sun regions and active regions to magnetically active stars, \cite{2003ApJ...598.1387P} demonstrated that the X-ray radiation follows a nearly universal power-law relation with the total unsigned magnetic flux,
\begin{eqnarray}
F_{\rm X}\propto \Phi^{\alpha},
\end{eqnarray}
where $F_{\rm X}$ denotes the X-ray flux and $\Phi$ is the total magnetic flux. They found a consistent power law index of $\alpha=1.15$ over more than 12 orders of magnitude in magnetic flux. This result suggests that the solar and stellar coronae are governed by a common magnetic heating process.

The power-law index $\alpha$ can be interpreted as a measure of the efficiency with which surface magnetic fields are converted into atmospheric heating and radiative losses. The superlinear behavior ($\alpha >1$) indicates that X-ray emission increases faster than magnetic flux, implying that more magnetically active domains heat the corona more efficiently.

\subsection{Solar Multi-Wavelength Scaling}

To investigate whether the magnetic-flux scaling extends beyond coronal X-ray emission, \citet{2022ApJ...927..179T} analyzed more than 10 years of Sun-as-a-star synoptic observations covering nearly an entire solar cycle. The study combined daily measurements of the total unsigned magnetic flux derived from full-disk SDO/HMI magnetograms with irradiance observations spanning a broad range of wavelengths from X-rays to radio bands, with the corresponding formation temperatures spanning from the corona to the chromosphere.

Figure \ref{fig:cc} presents representative examples of the resulting magnetic flux-irradiance relations for several atmospheric diagnostics. All diagnostics exhibit clear positive correlations with magnetic flux and can be described by power-law relations of the form $F\propto \Phi^{\alpha}$. The X-ray emission and coronal Fe XV line show nearly identical superlinear slopes ($\alpha\approx 1.15$), consistent with that in \citet{2003ApJ...598.1387P}. In contrast, the cooler lines such as C II, Ly$\alpha$, and Mg II k$+$h display sublinear slopes ($\alpha<1$). These results demonstrate that magnetic activity controls radiative output throughout the solar atmosphere, while the efficiency of magnetic heating varies systematically with the temperature.

\subsection{Comparison Between the Sun and Sun-like Stars}

To examine whether the solar scaling laws are applicable to cool stars, the relations derived from the solar observations were compared with measurements of Sun-like stars spanning a wide range of ages and activity levels. The stellar sample consists primarily of G-type main-sequence stars with ages ranging from approximately 50 Myr to 4.5 Gyr, including highly active young solar analogs such as EK Dra and $\kappa^{1}$ Cet and more evolved stars comparable to the present-day Sun \citep[e.g.][]{2020A&A...635A.142K}.

It is seen in Figure \ref{fig:cc} that, remarkably, the stellar data points lie on extensions of the solar scaling laws in all temperature regimes. This agreement is found not only for coronal emissions, where the universality of the magnetic flux-X-ray relation had previously been recognized, but also for transition-region and chromospheric lines.

These results provide strong observational evidence that a common magnetic-heating mechanism operates across the Sun and stars despite their vastly different ages and activity levels. The continuity between solar and stellar data suggests that the same underlying physical processes regulate atmospheric heating across the stars.

\subsection{Temperature Dependence of the Scaling Laws}

While the solar and stellar data follow common magnetic flux-irradiance scaling laws, the power-law index itself exhibits a strong dependence on atmospheric temperature. To investigate this behavior in greater detail, \citet{2022ApJS..262...46T} systematically analyzed a broad set of solar activity proxies and spectral irradiances and compiled a catalog of power-law indices covering atmospheric layers from the chromosphere to the corona.

\begin{figure}[t!]
	\centering
	\includegraphics[width=\linewidth]{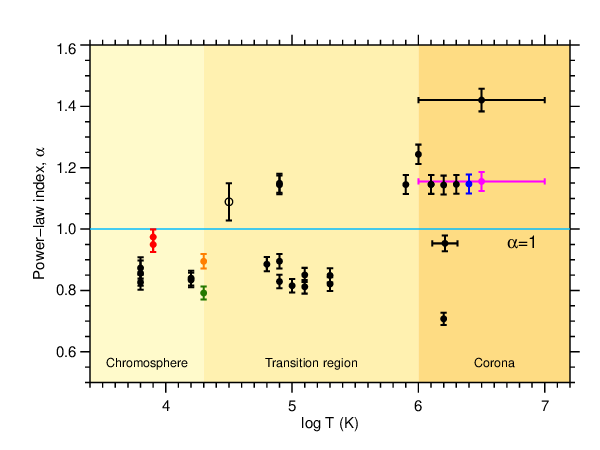}
	\caption{Power-law index $\alpha$ as a function of the formation temperature of spectral line. Coronal emissions generally exhibit superlinear scaling ($\alpha > 1$), whereas chromospheric and transition-region emissions show sublinear scaling ($\alpha < 1$). Figure is reproduced from \citet{2022ApJS..262...46T}.}
	\label{fig:pl}
\end{figure}

Figure \ref{fig:pl} summarizes the variation of the power-law index $\alpha$ as a function of the formation temperature of the corresponding spectral diagnostics. A clear temperature dependence is observed. Coronal emissions formed at temperatures above $\log{T}\sim 6$ generally exhibit superlinear relations with $\alpha>1$. In contrast, transition-region and chromospheric diagnostics show systematically smaller values, typically $\alpha < 1$, indicating a weaker response of the radiative output to increasing magnetic flux \citep[see, e.g.][]{1989ApJ...337..964S}.

These results indicate that the efficiency of magnetic heating varies throughout the atmosphere, becoming more sensitive to magnetic flux in the corona than in cooler layers. The observed temperature dependence provides an important constraint on models of solar and stellar atmospheric heating.

\section{Reconstruction of Stellar XUV Spectra}

\subsection{Motivation}

Direct observations of stellar extreme ultraviolet (EUV) radiation are challenging because interstellar hydrogen efficiently absorbs photons at these wavelengths. Nevertheless, the XUV radiation play a critical role in controlling the ionization, photochemistry, and escape of planetary atmospheres, making reliable estimates of stellar XUV spectra essential for exoplanet studies. The universal magnetic flux-irradiance scaling laws described in Section \ref{sec:scaling} provide a promising framework for predicting stellar high-energy radiation from observable magnetic properties.

\subsection{Empirical Reconstruction Method}

\begin{figure}[t!]
	\centering
	\includegraphics[width=0.85\linewidth]{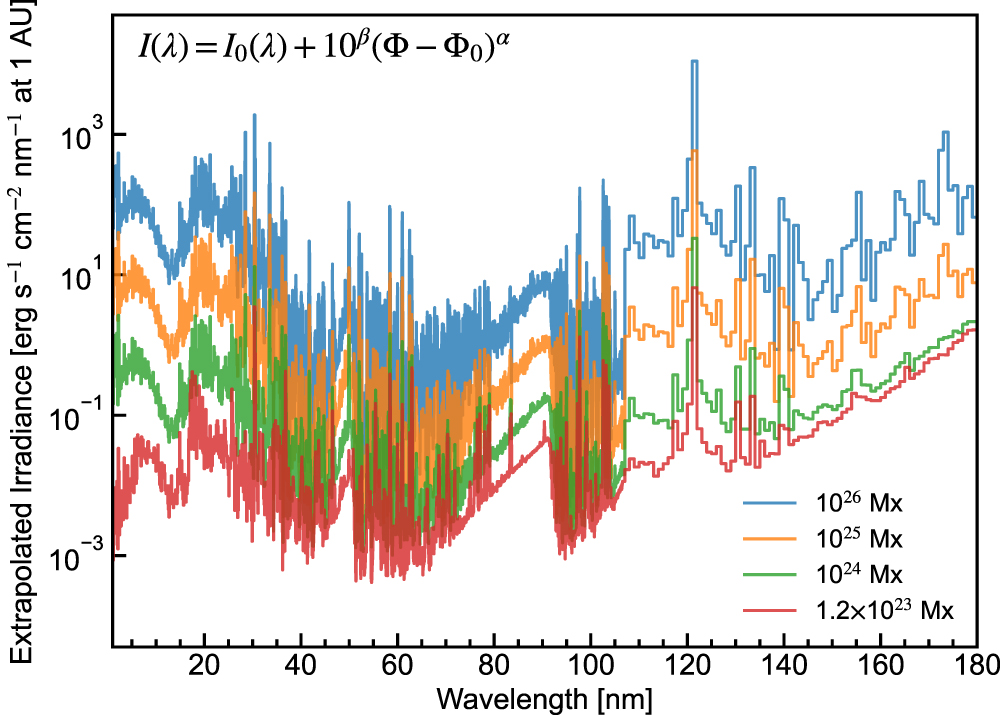}
	\caption{Sample synthesized spectra for stars having a total unsigned magnetic flux of $10^{24}$, $10^{25}$, and $10^{26}$ Mx. The model spectrum at a solar minimum value of $1.2 \times 10^{23}$ Mx is also plotted in red as a reference. Figure is reproduced from \citet{2023ApJ...945..147N}.}
	\label{fig:sample}
\end{figure}

Building on the solar-stellar scaling laws, \citet{2023ApJ...945..147N} developed an empirical method for reconstructing stellar XUV spectra from measured magnetic fluxes. Instead of deriving scaling laws for individual spectral lines, they established wavelength-dependent relations using solar observations spanning approximately 1–1800 {\AA}. The spectral irradiance at each wavelength was expressed as
\begin{eqnarray}
I(\lambda)=10^{\beta(\lambda)}\Phi^{\alpha(\lambda)},
\end{eqnarray}
where $I(\lambda)$ is the irradiance at wavelength $\lambda$, $\Phi$ is the total unsigned magnetic flux, and $\alpha(\lambda)$ and $\beta(\lambda)$ are empirically derived power-law coefficients. The sample
spectra for the stars with magnetic flux, $\Phi$, of $10^{23}$ to $10^{26}$ Mx are shown in Figure \ref{fig:sample}. The model spectra are available on GitHub.\footnote{\url{https://github.com/KosukeNamekata/StellarXUV.git}}

\subsection{Validation with Actual Stars}

The method was tested using several active solar analogs, including $\kappa^{1}$ Cet, $\pi^{1}$ UMa, and EK Dra. These stars are considerably younger and more active than the present-day Sun and therefore provide an important test of the applicability of the solar scaling laws.

\begin{figure*}[t!]
	\centering
	\includegraphics[width=140mm]{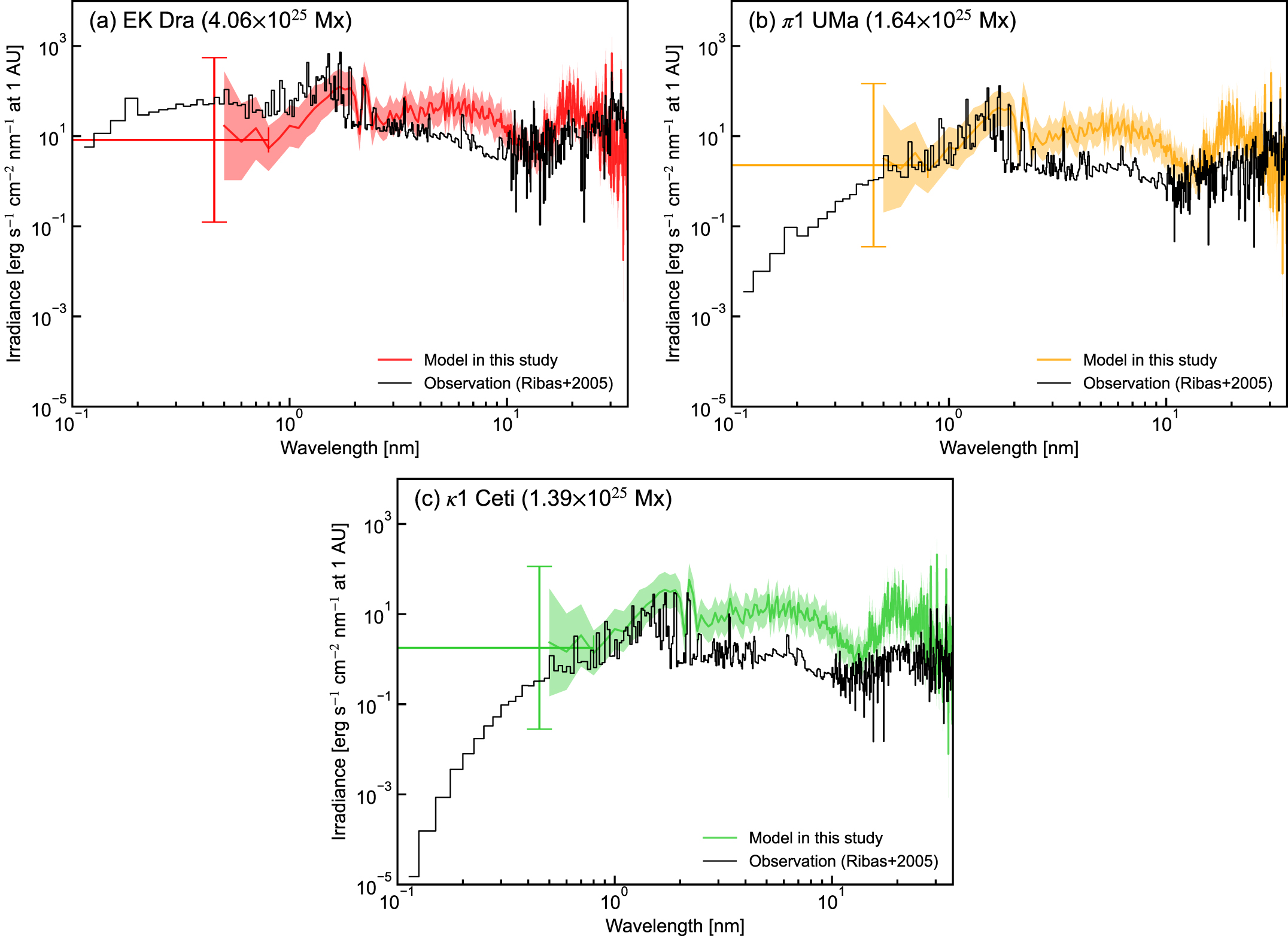}
	\caption{Comparison between observed and reconstructed XUV spectra for the young solar analogs (a) EK Dra, (b) $\pi^{1}$ UMa, and (c) $\kappa^{1}$ Cet. The empirical reconstructions reproduce the observed spectra over a broad wavelength range. Figure is reproduced from \citet{2023ApJ...945..147N}.}
	\label{fig:xuv}
\end{figure*}

Figure \ref{fig:xuv} compares the XUV spectra reconstructed from the observed magnetic fluxes (colored curves) with the corresponding observed spectra (black curves). The synthesized spectra successfully reproduce both the overall spectral shape and the observed flux levels over a broad wavelength range. The agreement is generally excellent for all targets. One notable exception is the X-ray range below approximately 1 {\AA} in EK Dra, where the reconstructed spectrum exceeds the observed flux. A possible explanation is that EK Dra produces frequent superflares, such that flare-related emission contributes to the quiescent coronal X-ray component and affects the observed spectral distribution.

\subsection{Implications}

The ability to reconstruct stellar XUV spectra from magnetic flux measurements provides a practical tool for characterizing the radiation environments of active stars. Because EUV observations are unavailable for most stars, the method offers a new approach for estimating stellar ionizing radiation and its impact on planetary atmospheres. Combined with the universal scaling laws discussed in Section \ref{sec:scaling}, these results support the idea that magnetic heating follows common physical principles across the Sun and cool stars and can be used to predict high-energy stellar radiation relevant to exoplanet habitability studies.

\section{Discussion}

The results summarized in the previous sections suggest that magnetic heating follows remarkably similar scaling laws across the Sun and Sun-like stars despite their wide range of ages and activity levels. The continuity between solar and stellar data implies that a common physical mechanism governs the conversion of magnetic energy into atmospheric radiation from the chromosphere to the corona. The agreement is particularly striking given that the stellar sample spans more than two orders of magnitude in age and magnetic activity.

At the same time, the temperature dependence of the power-law index $\alpha$ indicates that the response of atmospheric heating to magnetic activity is not uniform throughout the atmosphere. While superlinear relations ($\alpha>1$) are consistently observed in coronal emissions, chromospheric and transition-region lines exhibit sublinear scaling ($\alpha<1$). Existing theoretical and numerical studies have successfully reproduced the coronal behavior, but the physical origin of the weaker response in cooler atmospheres remains less well understood \citep[see, e.g.][]{1989ApJ...337..964S,2024NatAs...8..697B}. The observed temperature dependence therefore provides an important constraint on future models of solar and stellar atmospheric heating.

An important next step is to examine whether the same scaling laws remain valid for stars outside the parameter range explored so far. Preliminary studies suggest that pre-main-sequence and zero-age main-sequence stars may exhibit deviations from the solar-scaling relations, possibly reflecting differences in magnetic topology or dynamo processes \citep{2025ApJ...985...46Y}. Expanding the stellar sample and improving magnetic measurements will therefore be essential for better understanding the universality of magnetic heating across the broader cool-star population.

Future observations will provide new opportunities to investigate these questions. In particular, next-generation solar observatories capable of simultaneously probing atmospheric layers from the chromosphere to the corona, as represented by the EUV spectroscopic mission SOLAR-C \citep{2021SPIE11444E..0NS}, will enable more direct validations of the scaling laws and their underlying physical mechanisms. On the stellar side, missions such as Habitable Worlds Observatory, NewAthena \citep{2025NatAs...9...36C}, LAPYUTA \citep{2024SPIE13093E..0IT}, ESCAPE \citep{2026arXiv260800683Y} will be crucial in this effort. Such observations will strengthen the connection between solar and stellar physics and improve our understanding of the high-energy radiation environments surrounding cool stars.

\section{Summary}

The studies reviewed in this paper suggest that magnetic heating follows universal scaling laws across the Sun and cool stars. More than a decade of Sun-as-a-star observations reveals that radiative output from the chromosphere to the corona scales with the photospheric magnetic flux through power-law relations, although the heating efficiency depends on the temperature. Remarkably, G-type stars spanning ages from 50 Myr to 4.5 Gyr follow extensions of the solar scaling laws, providing strong evidence for a common magnetic-heating mechanism operating across a broad range of stellar activity levels. These empirical relations further enable the reconstruction of stellar XUV spectra from magnetic flux measurements, offering a practical tool for characterizing the radiation environments of active stars. Future observations and numerical modeling will help clarify the physical origin of the observed temperature dependence and further strengthen the solar-stellar connection.

\section*{Acknowledgments}
{This work was supported by JSPS KAKENHI
grant No. JP25K01041 (PI: K. Namekata) and NASA Living With A Star Program 80NSSC26K0039.}

\bibliographystyle{cs23proc}
\bibliography{example.bib}

@ARTICLE{2020IJAsB..19..136A,
       author = {{Airapetian}, V.~S. and {Barnes}, R. and {Cohen}, O. and {Collinson}, G.~A. and {Danchi}, W.~C. and {Dong}, C.~F. and {Del Genio}, A.~D. and {France}, K. and {Garcia-Sage}, K. and {Glocer}, A. and {Gopalswamy}, N. and {Grenfell}, J.~L. and {Gronoff}, G. and {G{\"u}del}, M. and {Herbst}, K. and {Henning}, W.~G. and {Jackman}, C.~H. and {Jin}, M. and {Johnstone}, C.~P. and {Kaltenegger}, L. and {Kay}, C.~D. and {Kobayashi}, K. and {Kuang}, W. and {Li}, G. and {Lynch}, B.~J. and {L{\"u}ftinger}, T. and {Luhmann}, J.~G. and {Maehara}, H. and {Mlynczak}, M.~G. and {Notsu}, Y. and {Osten}, R.~A. and {Ramirez}, R.~M. and {Rugheimer}, S. and {Scheucher}, M. and {Schlieder}, J.~E. and {Shibata}, K. and {Sousa-Silva}, C. and {Stamenkovi{\'c}}, V. and {Strangeway}, R.~J. and {Usmanov}, A.~V. and {Vergados}, P. and {Verkhoglyadova}, O.~P. and {Vidotto}, A.~A. and {Voytek}, M. and {Way}, M.~J. and {Zank}, G.~P. and {Yamashiki}, Y.},
        title = "{Impact of space weather on climate and habitability of terrestrial-type exoplanets}",
      journal = {International Journal of Astrobiology},
         year = 2020,
        month = apr,
       volume = {19},
       number = {2},
        pages = {136-194},
          doi = {10.1017/S1473550419000132},
archivePrefix = {arXiv},
       eprint = {1905.05093},
 primaryClass = {astro-ph.EP},
       adsurl = {https://ui.adsabs.harvard.edu/abs/2020IJAsB..19..136A}
}

@ARTICLE{2016NatGe...9..452A,
       author = {{Airapetian}, V.~S. and {Glocer}, A. and {Gronoff}, G. and {H{\'e}brard}, E. and {Danchi}, W.},
        title = "{Prebiotic chemistry and atmospheric warming of early Earth by an active young Sun}",
      journal = {Nature Geoscience},
         year = 2016,
        month = jun,
       volume = {9},
       number = {6},
        pages = {452-455},
          doi = {10.1038/ngeo2719},
       adsurl = {https://ui.adsabs.harvard.edu/abs/2016NatGe...9..452A}
}

@ARTICLE{2024NatAs...8..697B,
       author = {{Bose}, Souvik and {De Pontieu}, Bart and {Hansteen}, Viggo and {Sainz Dalda}, Alberto and {Savage}, Sabrina and {Winebarger}, Amy},
        title = "{Chromospheric and coronal heating in an active region plage by dissipation of currents from braiding}",
      journal = {Nature Astronomy},
         year = 2024,
        month = jun,
       volume = {8},
        pages = {697-705},
          doi = {10.1038/s41550-024-02241-8},
archivePrefix = {arXiv},
       eprint = {2211.08579},
 primaryClass = {astro-ph.SR},
       adsurl = {https://ui.adsabs.harvard.edu/abs/2024NatAs...8..697B}
}

@ARTICLE{2017LRSP...14....4B,
       author = {{Brun}, Allan Sacha and {Browning}, Matthew K.},
        title = "{Magnetism, dynamo action and the solar-stellar connection}",
      journal = {Living Reviews in Solar Physics},
         year = 2017,
        month = dec,
       volume = {14},
       number = {1},
          eid = {4},
        pages = {4},
          doi = {10.1007/s41116-017-0007-8},
       adsurl = {https://ui.adsabs.harvard.edu/abs/2017LRSP...14....4B}
}

@ARTICLE{2025NatAs...9...36C,
       author = {{Cruise}, Mike and {Guainazzi}, Matteo and {Aird}, James and {Carrera}, Francisco J. and {Costantini}, Elisa and {Corrales}, Lia and {Dauser}, Thomas and {Eckert}, Dominique and {Gastaldello}, Fabio and {Matsumoto}, Hironori and {Osten}, Rachel and {Petrucci}, Pierre-Olivier and {Porquet}, Delphine and {Pratt}, Gabriel W. and {Rea}, Nanda and {Reiprich}, Thomas H. and {Simionescu}, Aurora and {Spiga}, Daniele and {Troja}, Eleonora},
        title = "{The NewAthena mission concept in the context of the next decade of X-ray astronomy}",
      journal = {Nature Astronomy},
         year = 2025,
        month = jan,
       volume = {9},
        pages = {36-44},
          doi = {10.1038/s41550-024-02416-3},
archivePrefix = {arXiv},
       eprint = {2501.03100},
 primaryClass = {astro-ph.IM},
       adsurl = {https://ui.adsabs.harvard.edu/abs/2025NatAs...9...36C}
}

@ARTICLE{1998ApJ...508..885F,
       author = {{Fisher}, George H. and {Longcope}, Dana W. and {Metcalf}, Thomas R. and {Pevtsov}, Alexei A.},
        title = "{Coronal Heating in Active Regions as a Function of Global Magnetic Variables}",
      journal = {\apj},
         year = 1998,
        month = dec,
       volume = {508},
       number = {2},
        pages = {885-898},
          doi = {10.1086/306435},
       adsurl = {https://ui.adsabs.harvard.edu/abs/1998ApJ...508..885F}
}

@ARTICLE{2004A&ARv..12...71G,
       author = {{G{\"u}del}, Manuel},
        title = "{X-ray astronomy of stellar coronae}",
      journal = {\aapr},
         year = 2004,
        month = sep,
       volume = {12},
       number = {2-3},
        pages = {71-237},
          doi = {10.1007/s00159-004-0023-2},
archivePrefix = {arXiv},
       eprint = {astro-ph/0406661},
 primaryClass = {astro-ph},
       adsurl = {https://ui.adsabs.harvard.edu/abs/2004A&ARv..12...71G}
}

@ARTICLE{2006SoPh..234...41K,
       author = {{Klimchuk}, James A.},
        title = "{On Solving the Coronal Heating Problem}",
      journal = {\solphys},
         year = 2006,
        month = mar,
       volume = {234},
       number = {1},
        pages = {41-77},
          doi = {10.1007/s11207-006-0055-z},
archivePrefix = {arXiv},
       eprint = {astro-ph/0511841},
 primaryClass = {astro-ph},
       adsurl = {https://ui.adsabs.harvard.edu/abs/2006SoPh..234...41K}
}

@ARTICLE{2020A&A...635A.142K,
       author = {{Kochukhov}, O. and {Hackman}, T. and {Lehtinen}, J.~J. and {Wehrhahn}, A.},
        title = "{Hidden magnetic fields of young suns}",
      journal = {\aap},
         year = 2020,
        month = mar,
       volume = {635},
          eid = {A142},
        pages = {A142},
          doi = {10.1051/0004-6361/201937185},
archivePrefix = {arXiv},
       eprint = {2002.10469},
 primaryClass = {astro-ph.SR},
       adsurl = {https://ui.adsabs.harvard.edu/abs/2020A&A...635A.142K}
}

@BOOK{2019LNP...955.....L,
       author = {{Linsky}, Jeffrey},
        title = "{Host Stars and their Effects on Exoplanet Atmospheres}",
         year = 2019,
       volume = {955},
          doi = {10.1007/978-3-030-11452-7},
       adsurl = {https://ui.adsabs.harvard.edu/abs/2019LNP...955.....L}
}

@ARTICLE{2012Natur.485..478M,
       author = {{Maehara}, Hiroyuki and {Shibayama}, Takuya and {Notsu}, Shota and {Notsu}, Yuta and {Nagao}, Takashi and {Kusaba}, Satoshi and {Honda}, Satoshi and {Nogami}, Daisaku and {Shibata}, Kazunari},
        title = "{Superflares on solar-type stars}",
      journal = {\nat},
         year = 2012,
        month = may,
       volume = {485},
       number = {7399},
        pages = {478-481},
          doi = {10.1038/nature11063},
       adsurl = {https://ui.adsabs.harvard.edu/abs/2012Natur.485..478M}
}

@ARTICLE{2022NatAs...6..241N,
       author = {{Namekata}, Kosuke and {Maehara}, Hiroyuki and {Honda}, Satoshi and {Notsu}, Yuta and {Okamoto}, Soshi and {Takahashi}, Jun and {Takayama}, Masaki and {Ohshima}, Tomohito and {Saito}, Tomoki and {Katoh}, Noriyuki and {Tozuka}, Miyako and {Murata}, Katsuhiro L. and {Ogawa}, Futa and {Niwano}, Masafumi and {Adachi}, Ryo and {Oeda}, Motoki and {Shiraishi}, Kazuki and {Isogai}, Keisuke and {Seki}, Daikichi and {Ishii}, Takako T. and {Ichimoto}, Kiyoshi and {Nogami}, Daisaku and {Shibata}, Kazunari},
        title = "{Probable detection of an eruptive filament from a superflare on a solar-type star}",
      journal = {Nature Astronomy},
         year = 2021,
        month = dec,
       volume = {6},
        pages = {241-248},
          doi = {10.1038/s41550-021-01532-8},
archivePrefix = {arXiv},
       eprint = {2112.04808},
 primaryClass = {astro-ph.SR},
       adsurl = {https://ui.adsabs.harvard.edu/abs/2022NatAs...6..241N}
}

@ARTICLE{2023ApJ...945..147N,
       author = {{Namekata}, Kosuke and {Toriumi}, Shin and {Airapetian}, Vladimir S. and {Shoda}, Munehito and {Watanabe}, Kyoko and {Notsu}, Yuta},
        title = "{Reconstructing the XUV Spectra of Active Sun-like Stars Using Solar Scaling Relations with Magnetic Flux}",
      journal = {\apj},
         year = 2023,
        month = mar,
       volume = {945},
       number = {2},
          eid = {147},
        pages = {147},
          doi = {10.3847/1538-4357/acbe38},
archivePrefix = {arXiv},
       eprint = {2302.10376},
 primaryClass = {astro-ph.SR},
       adsurl = {https://ui.adsabs.harvard.edu/abs/2023ApJ...945..147N}
}

@ARTICLE{2013ApJ...771..127N,
       author = {{Notsu}, Yuta and {Shibayama}, Takuya and {Maehara}, Hiroyuki and {Notsu}, Shota and {Nagao}, Takashi and {Honda}, Satoshi and {Ishii}, Takako T. and {Nogami}, Daisaku and {Shibata}, Kazunari},
        title = "{Superflares on Solar-type Stars Observed with Kepler II. Photometric Variability of Superflare-generating Stars: A Signature of Stellar Rotation and Starspots}",
      journal = {\apj},
         year = 2013,
        month = jul,
       volume = {771},
       number = {2},
          eid = {127},
        pages = {127},
          doi = {10.1088/0004-637X/771/2/127},
archivePrefix = {arXiv},
       eprint = {1304.7361},
 primaryClass = {astro-ph.SR},
       adsurl = {https://ui.adsabs.harvard.edu/abs/2013ApJ...771..127N}
}

@ARTICLE{1984ApJ...279..763N,
       author = {{Noyes}, R.~W. and {Hartmann}, L.~W. and {Baliunas}, S.~L. and {Duncan}, D.~K. and {Vaughan}, A.~H.},
        title = "{Rotation, convection, and magnetic activity in lower main-sequence stars.}",
      journal = {\apj},
         year = 1984,
        month = apr,
       volume = {279},
        pages = {763-777},
          doi = {10.1086/161945},
       adsurl = {https://ui.adsabs.harvard.edu/abs/1984ApJ...279..763N}
}

@ARTICLE{2003ApJ...598.1387P,
       author = {{Pevtsov}, Alexei A. and {Fisher}, George H. and {Acton}, Loren W. and {Longcope}, Dana W. and {Johns-Krull}, Christopher M. and {Kankelborg}, Charles C. and {Metcalf}, Thomas R.},
        title = "{The Relationship Between X-Ray Radiance and Magnetic Flux}",
      journal = {\apj},
         year = 2003,
        month = dec,
       volume = {598},
       number = {2},
        pages = {1387-1391},
          doi = {10.1086/378944},
       adsurl = {https://ui.adsabs.harvard.edu/abs/2003ApJ...598.1387P}
}

@ARTICLE{2005ApJ...622..680R,
       author = {{Ribas}, Ignasi and {Guinan}, Edward F. and {G{\"u}del}, Manuel and {Audard}, Marc},
        title = "{Evolution of the Solar Activity over Time and Effects on Planetary Atmospheres. I. High-Energy Irradiances (1-1700 {\r{A}})}",
      journal = {\apj},
         year = 2005,
        month = mar,
       volume = {622},
       number = {1},
        pages = {680-694},
          doi = {10.1086/427977},
archivePrefix = {arXiv},
       eprint = {astro-ph/0412253},
 primaryClass = {astro-ph},
       adsurl = {https://ui.adsabs.harvard.edu/abs/2005ApJ...622..680R}
}

@ARTICLE{1989ApJ...337..964S,
       author = {{Schrijver}, C.~J. and {Cote}, J. and {Zwaan}, C. and {Saar}, S.~H.},
        title = "{Relations between the Photospheric Magnetic Field and the Emission from the Outer Atmospheres of Cool Stars. I. The Solar CA II K Line Core Emission}",
      journal = {\apj},
         year = 1989,
        month = feb,
       volume = {337},
        pages = {964},
          doi = {10.1086/167168},
       adsurl = {https://ui.adsabs.harvard.edu/abs/1989ApJ...337..964S}
}

@ARTICLE{2011LRSP....8....6S,
       author = {{Shibata}, Kazunari and {Magara}, Tetsuya},
        title = "{Solar Flares: Magnetohydrodynamic Processes}",
      journal = {Living Reviews in Solar Physics},
         year = 2011,
        month = dec,
       volume = {8},
       number = {1},
          eid = {6},
        pages = {6},
          doi = {10.12942/lrsp-2011-6},
       adsurl = {https://ui.adsabs.harvard.edu/abs/2011LRSP....8....6S}
}

@INPROCEEDINGS{2021SPIE11444E..0NS,
       author = {{Shimizu}, Toshifumi and {Imada}, Shinsuke and {Kawate}, Tomoko and {Suematsu}, Yoshinori and {Hara}, Hirohisa and {Tsuzuki}, Toshihiro and {Katsukawa}, Yukio and {Kubo}, Masahito and {Ishikawa}, Ryoko and {Watanabe}, Tetsuya and {Toriumi}, Shin and {Ichimoto}, Kiyoshi and {Nagata}, Shin'ichi and {Hasegawa}, Takahiro and {Yokoyama}, Takaaki and {Watanabe}, Kyoko and {Tsuno}, Katsuhiko and {Korendyke}, Clarence M. and {Warren}, Harry and {De Pontieu}, Bart and {Boerner}, Paul and {Solanki}, Sami K. and {Teriaca}, Luca and {Schuehle}, Udo and {Matthews}, Sarah and {Long}, David and {Thomas}, William and {Hancock}, Barry and {Reid}, Hamish and {Fludra}, Andrzej and {Auch{\`e}re}, Frederic and {Andretta}, Vincenzo and {Naletto}, Giampiero and {Poletto}, Luca and {Harra}, Louise},
        title = "{The Solar-C (EUVST) mission: the latest status}",
    booktitle = {Society of Photo-Optical Instrumentation Engineers (SPIE) Conference Series},
         year = 2021,
       editor = {{den Herder}, Jan-Willem A. and {Nikzad}, Shouleh and {Nakazawa}, Kazuhiro},
       series = {Society of Photo-Optical Instrumentation Engineers (SPIE) Conference Series},
       volume = {11444},
        month = jan,
          eid = {114440N},
        pages = {114440N},
          doi = {10.1117/12.2560887},
       adsurl = {https://ui.adsabs.harvard.edu/abs/2021SPIE11444E..0NS}
}

@ARTICLE{1972ApJ...171..565S,
       author = {{Skumanich}, A.},
        title = "{Time Scales for Ca II Emission Decay, Rotational Braking, and Lithium Depletion}",
      journal = {\apj},
         year = 1972,
        month = feb,
       volume = {171},
        pages = {565},
          doi = {10.1086/151310},
       adsurl = {https://ui.adsabs.harvard.edu/abs/1972ApJ...171..565S}
}

@ARTICLE{2006JGRA..111.6101S,
       author = {{Suzuki}, Takeru K. and {Inutsuka}, Shu-Ichiro},
        title = "{Solar winds driven by nonlinear low-frequency Alfv{\'e}n waves from the photosphere: Parametric study for fast/slow winds and disappearance of solar winds}",
      journal = {Journal of Geophysical Research (Space Physics)},
         year = 2006,
        month = jun,
       volume = {111},
       number = {A6},
          eid = {A06101},
        pages = {A06101},
          doi = {10.1029/2005JA011502},
archivePrefix = {arXiv},
       eprint = {astro-ph/0511006},
 primaryClass = {astro-ph},
       adsurl = {https://ui.adsabs.harvard.edu/abs/2006JGRA..111.6101S}
}

@ARTICLE{2026SoPh..301...48T,
       author = {{Toriumi}, Shin},
        title = "{Bridging Solar and Stellar Physics: Role of SDO in Understanding Stellar Active Regions and Atmospheric Heating}",
      journal = {\solphys},
         year = 2026,
        month = mar,
       volume = {301},
       number = {3},
          eid = {48},
        pages = {48},
          doi = {10.1007/s11207-026-02634-0},
archivePrefix = {arXiv},
       eprint = {2602.16779},
 primaryClass = {astro-ph.SR},
       adsurl = {https://ui.adsabs.harvard.edu/abs/2026SoPh..301...48T}
}

@ARTICLE{2022ApJ...927..179T,
       author = {{Toriumi}, Shin and {Airapetian}, Vladimir S.},
        title = "{Universal Scaling Laws for Solar and Stellar Atmospheric Heating}",
      journal = {\apj},
         year = 2022,
        month = mar,
       volume = {927},
       number = {2},
          eid = {179},
        pages = {179},
          doi = {10.3847/1538-4357/ac5179},
archivePrefix = {arXiv},
       eprint = {2202.01232},
 primaryClass = {astro-ph.SR},
       adsurl = {https://ui.adsabs.harvard.edu/abs/2022ApJ...927..179T}
}

@ARTICLE{2022ApJS..262...46T,
       author = {{Toriumi}, Shin and {Airapetian}, Vladimir S. and {Namekata}, Kosuke and {Notsu}, Yuta},
        title = "{Universal Scaling Laws for Solar and Stellar Atmospheric Heating: Catalog of Power-law Index between Solar Activity Proxies and Various Spectral Irradiances}",
      journal = {\apjs},
         year = 2022,
        month = oct,
       volume = {262},
       number = {2},
          eid = {46},
        pages = {46},
          doi = {10.3847/1538-4365/ac8b15},
archivePrefix = {arXiv},
       eprint = {2208.10511},
 primaryClass = {astro-ph.SR},
       adsurl = {https://ui.adsabs.harvard.edu/abs/2022ApJS..262...46T}
}

@ARTICLE{2017ApJ...834...56T,
       author = {{Toriumi}, Shin and {Schrijver}, Carolus J. and {Harra}, Louise K. and {Hudson}, Hugh and {Nagashima}, Kaori},
        title = "{Magnetic Properties of Solar Active Regions That Govern Large Solar Flares and Eruptions}",
      journal = {\apj},
         year = 2017,
        month = jan,
       volume = {834},
       number = {1},
          eid = {56},
        pages = {56},
          doi = {10.3847/1538-4357/834/1/56},
archivePrefix = {arXiv},
       eprint = {1611.05047},
 primaryClass = {astro-ph.SR},
       adsurl = {https://ui.adsabs.harvard.edu/abs/2017ApJ...834...56T}
}

@ARTICLE{2019LRSP...16....3T,
       author = {{Toriumi}, Shin and {Wang}, Haimin},
        title = "{Flare-productive active regions}",
      journal = {Living Reviews in Solar Physics},
         year = 2019,
        month = dec,
       volume = {16},
       number = {1},
          eid = {3},
        pages = {3},
          doi = {10.1007/s41116-019-0019-7},
archivePrefix = {arXiv},
       eprint = {1904.12027},
 primaryClass = {astro-ph.SR},
       adsurl = {https://ui.adsabs.harvard.edu/abs/2019LRSP...16....3T}
}

@INPROCEEDINGS{2024SPIE13093E..0IT,
       author = {{Tsuchiya}, Fuminori and {Murakami}, Go and {Yamazaki}, Atsushi and {Kameda}, Shingo and {Kimura}, Tomoki and {Koga}, Ryoichi and {Masunaga}, Kei and {Sakai}, Shotaro and {Ikoma}, Masahiro and {Nakayama}, Akifumi and {Ouchi}, Masami and {Tanaka}, Masaomi and {Toriumi}, Shin and {Kagitani}, Masato and {Yoshioka}, Kazuo and {Tao}, Chihiro and {Kita}, Hajime and {Yajima}, Hidenobu and {Sagawa}, Hideo and {Nakagawa}, Hiromu and {Hamori}, Hitoshi and {Kimura}, Jun and {Enya}, Keigo and {Namekata}, Kosuke and {Yamada}, Manabu and {Kuwabara}, Masaki and {Terada}, Naoki and {Ozaki}, Naoya and {Narita}, Norio and {Aizawa}, Sae and {Takagi}, Seiko and {Sakai}, Shinitiro and {Aoki}, Shohei and {Matsuda}, Shoya and {Tan}, Shuya and {Sumi}, Takahiro and {Kodama}, Takanori and {Moriya}, Takashi and {Shibuya}, Takatoshi and {Satoh}, Takehiko and {Kawano}, Taro and {Tominaga}, Nozomu and {Shimizu}, Toshifumi and {Kasaba}, Yasumasa and {Yatsu}, Yoichi and {Ono}, Yoshiaki and {Suzuki}, Yudai and {Matsuda}, Yuichi and {Harada}, Yuki and {Notsu}, Yuta},
        title = "{Overview of the LAPYUTA mission (Life-environmentology, Astronomy, and PlanetarY Ultraviolet Telescope Assembly)}",
    booktitle = {Space Telescopes and Instrumentation 2024: Ultraviolet to Gamma Ray},
         year = 2024,
       editor = {{den Herder}, Jan-Willem A. and {Nikzad}, Shouleh and {Nakazawa}, Kazuhiro},
       series = {Society of Photo-Optical Instrumentation Engineers (SPIE) Conference Series},
       volume = {13093},
        month = aug,
          eid = {130930I},
        pages = {130930I},
          doi = {10.1117/12.3017298},
       adsurl = {https://ui.adsabs.harvard.edu/abs/2024SPIE13093E..0IT}
}

@ARTICLE{2014MNRAS.441.2361V,
       author = {{Vidotto}, A.~A. and {Gregory}, S.~G. and {Jardine}, M. and {Donati}, J.~F. and {Petit}, P. and {Morin}, J. and {Folsom}, C.~P. and {Bouvier}, J. and {Cameron}, A.~C. and {Hussain}, G. and {Marsden}, S. and {Waite}, I.~A. and {Fares}, R. and {Jeffers}, S. and {do Nascimento}, J.~D.},
        title = "{Stellar magnetism: empirical trends with age and rotation}",
      journal = {\mnras},
         year = 2014,
        month = jul,
       volume = {441},
       number = {3},
        pages = {2361-2374},
          doi = {10.1093/mnras/stu728},
archivePrefix = {arXiv},
       eprint = {1404.2733},
 primaryClass = {astro-ph.SR},
       adsurl = {https://ui.adsabs.harvard.edu/abs/2014MNRAS.441.2361V}
}

@ARTICLE{2011ApJ...743...48W,
       author = {{Wright}, Nicholas J. and {Drake}, Jeremy J. and {Mamajek}, Eric E. and {Henry}, Gregory W.},
        title = "{The Stellar-activity-Rotation Relationship and the Evolution of Stellar Dynamos}",
      journal = {\apj},
         year = 2011,
        month = dec,
       volume = {743},
       number = {1},
          eid = {48},
        pages = {48},
          doi = {10.1088/0004-637X/743/1/48},
archivePrefix = {arXiv},
       eprint = {1109.4634},
 primaryClass = {astro-ph.SR},
       adsurl = {https://ui.adsabs.harvard.edu/abs/2011ApJ...743...48W}
}

@ARTICLE{2025ApJ...985...46Y,
       author = {{Yamashita}, Mai and {Itoh}, Yoichi and {Toriumi}, Shin},
        title = "{Variations in the Magnetic Field Strength of Pre-main-sequence Stars, Solar-type Main-sequence Stars, and the Sun}",
      journal = {\apj},
         year = 2025,
        month = may,
       volume = {985},
       number = {1},
          eid = {46},
        pages = {46},
          doi = {10.3847/1538-4357/adc816},
archivePrefix = {arXiv},
       eprint = {2504.04684},
 primaryClass = {astro-ph.SR},
       adsurl = {https://ui.adsabs.harvard.edu/abs/2025ApJ...985...46Y}
}

\end{document}